\documentstyle[12pt,amssymb,amsmath]{article}
\def\text#1{\mbox{#1}}

\begin{document}

\title{{\bf Nonabelian topological mass mechanism for a three-dimensional
2-form field }}

\author{
{\large {\bf D. M. Medeiros$^{1,2}$, R. R. Landim$^{1}$, and C. A.
S. Almeida$^{1}$ \thanks{ E-mail address: carlos@fisica.ufc.br} }}
\\ \\
{\normalsize {\it $^{1}$Universidade Federal do Cear\'{a} -
Departamento de F\'{\i}sica }}
\\
{\normalsize {\it C.P. 6030, 60470-455 Fortaleza-Ce , Brazil }}
\\ \\
{\normalsize {\it $^{2}$Universidade Estadual do Cear\'{a}
- Departamento de F\'{\i}sica e Qu\'{\i}mica }}
\\
{\normalsize {\it Av. Paranjana, 1700, 60740-000 Fortaleza-Ce,
Brazil }} }

\maketitle

\begin{abstract}
Starting from a recently proposed abelian topological model in
(2+1) dimensions,  we use the method of the consistent
deformations to prove that a  topologically massive model
involving the Kalb-Ramond two form field does not admit a
nonabelian generalization. The introduction of a connection-type
one form field keeps the previous result. However we show that the
goal is achieved if we introduce a vectorial auxiliary field,
exhibiting a nonabelian topological mass generation mechanism in
$D=3$, that provides mass for the Kalb-Ramond field. Further, we
find the complete set of BRST and anti-BRST equations using the
horizontality condition, suggesting a connection between this
formalism and the method of the consistent deformations.

\end{abstract}

\medskip
PACS\, 11.15.-q, 11.10.-z,12.90.+b

\section{Introduction}

Antisymmetric tensor gauge fields provide a natural extension of
the usual vector gauge fields, appearing as mediator of string
interaction and having an important key role in supergravity.
Also, they are fundamental to the well known topological mass
generation mechanism \cite{jackiw} for abelian vector boson in
four dimensions, through a BF term~\cite{la1}. This term is
characterized by the presence of an antisymmetric gauge field
$B_{\mu \nu }$ ( Kalb-Ramond field) and the field strength $F_{\mu
\nu }.$ Nonabelian extensions of models involving antisymmetric
gauge fields in four dimensional space-time were introduced by
Hwang and Lee \cite{hwang} and Lahiri \cite{lahiri}, in the
context of topologically mass generation models. Both procedures
requires the introduction of an auxiliary vector field, justified
by the need to untie the constraint between two and three form
curvatures $F$ and $H$. A nonabelian theory involving an
antisymmetric tensor field coupled to a gauge field appears as an
alternative mechanism for generating vector bosons masses, similar
to the theory of a heavy Higgs particle \cite{lahirione}. It is
worth to mention a generalization to a compact nonabelian gauge
group of an abelian mechanism in the context of nonabelian quantum
hair on black holes \cite{lahiritwo}.

Kalb-Ramond fields arise naturally in string coupled to the area
element of the two-dimensional worldsheet \cite{KR} and a string
Higgs mechanism was introduced by Rey in ref. \cite{rey}.

On the other hand, using the technique of consistent deformation,
Henneaux {\it et al.}~\cite{no-go}, have proved that is not
possible to generalize the topological mass mechanism pointed out
above to its nonabelian counterpart with the same field contents
and fulfilling the power-counting renormalization requirements. In
this way, they put in more rigorous grounds the need to add an
auxiliary field.

Recently, we have shown a topological mass generation in an
abelian three-dimensional model involving a two-form gauge field
$B_{\mu \nu }$ and
a scalar field $\varphi $, rather than the usual Maxwell-Chern-Simons model~%
\cite{non-chern}. Also we have proved the classical duality
between a massless scalar field and a vector gauge field. The
action for the model just mentioned reads as

\begin{equation}
S_{inv}^A=\int \!\!d^3x~\left( \frac 1{12}H_{\mu \nu \alpha
}H^{\mu \nu \alpha }+\frac 12\partial _\mu \varphi \partial ^\mu
\varphi +\frac m2\epsilon ^{\mu \nu \alpha }B_{\mu \nu }\partial
_\alpha \varphi \right) , \label{sa}
\end{equation}
where $H_{\mu \nu \alpha }$ is the totally antisymmetric tensor
\begin{equation}
H_{\mu \nu \alpha }=\partial _\mu B_{\nu \alpha }+\partial _\alpha
B_{\mu \nu }+\partial _\nu B_{\alpha \mu }\text{ .}  \label{H}
\end{equation}
The action (\ref{sa}), is invariant under the transformation
\begin{equation}
\delta \varphi =0,\quad \quad \delta B_{\mu \nu }=\partial _{[\mu
}\omega _{\nu ]}\text{ ,}  \label{invsa}
\end{equation}
and its equations of motion give the massive equations

\begin{equation}
(\square +m^2)\partial _\mu \varphi =0  \label{gmt061}
\end{equation}
and
\begin{equation}
(\square +m^2)H_{\mu \nu \alpha }=0\text{ }.  \label{gmt062}
\end{equation}

The model described by action (\ref{sa}) can be consistently
obtained by dimensional reduction of a four-dimensional $B\wedge
F$ model if we discard the Chern-Simons-like terms
\cite{non-chern}.

The purpose of the present work is to construct a nonabelian
version of the action (\ref{sa}). We begin making an analysis of
the possibility to
construct the nonabelian action with $\varphi \rightarrow \varphi ^a$ and $%
B_{\mu \nu }\rightarrow B_{\mu \nu }^a$, i.e, with the same field
content, and the same number of local symmetries, by making use of
the method of consistent deformation~\cite{def}. As will be
proved, there is a no-go theorem for this construction. The same
occur with an introduction of a connection-type one-form gauge
field. The only possibility is via an introduction of an auxiliary
vector field. This introduction is made and we show a nonabelian
topological mass generation mechanism for the Kalb-Ramond field in
three dimensions.

This paper is organized as follows. In section 2 we apply the
method of consistent deformations to an abelian topological
three-dimensional model involving a two-form gauge field $B_{\mu
\nu }$ and a scalar field $\varphi $ in order to study possible
nonabelian generalizations. Then, a no-go theorem is established.
In section 3 we obtain the BRST and anti-BRST equations by
applying the horizontality condition, including an auxiliary
vectorial field, which allows the sought nonabelian
generalization. Section 4 presents a nonabelian topological mass
generation mechanism and finally we draw our conclusions in
section 5.

\section{Deforming consistently the abelian model}

Let us now apply the consistent deformation method described in
~\cite{def}. We shall start therefore with the following invariant
action
\begin{equation}
S_0^{\prime }=\int \!\!d^3x~\left( \frac 1{12}H_{\mu \nu \alpha
}^aH^{a\mu \nu \alpha }+\frac 12\partial _\mu \varphi ^a\partial
^\mu \varphi ^a\right) ,  \label{s0l}
\end{equation}
where now $\varphi ^a$ and $H_{\mu \nu \alpha }^a$ are scalar
fields and the abelian curvature tensor~(\ref{H}) for a set of $N$
fields. All fields are valued in the Lie algebra ${\cal G}$ of
some Lie group G. Since we are interested if the mass term can
exist in abelian extension of ~(\ref{sa}), the mass parameter will
be considered as a deformation parameter. The action ~(\ref{s0l})
is invariant under the transformations
\begin{equation}
\delta \varphi ^a=0~,\quad \quad \delta B_{\mu \nu }^a=\partial
_\mu \omega _\nu ^a-\partial _\nu \omega _\mu ^a.  \label{invs0l}
\end{equation}
Since the transformation of $B_{\mu \nu }^a$ is reducible, we
introduce a set of ghosts $(\eta _\mu ^a,\rho ^a)$, where $\eta
_\mu ^a$ is a ghost for the gauge transformation of $B_{\mu \nu
}^a$, and $\rho ^a$ the ghost for ghost for taking into account
this reducibility. For all fields of the model we introduce the
corresponding antifields $(B_{\mu \nu }^{*a},\varphi ^{*a},\eta
_\mu ^{*a},\rho ^{*a})$. The antifields action reads
\begin{equation}
S_{ant}^{\prime }=\int \!\!d^3x~\left( \frac 12B^{*\mu \nu
a}\partial _{[\mu }\eta _{\nu ]}^a+\eta ^{*\mu a}\partial _\mu
\rho ^a\right) .  \label{s0la}
\end{equation}
The free action
\begin{equation}
S_0=S_0^{\prime }+S_{ant}^{\prime },  \label{s0}
\end{equation}
is solution of the master equation
\begin{equation}
(S_0,S_0)=0,  \label{me}
\end{equation}
with
\begin{equation}
(S_0,S_0)=\int \!\!d^3x~\left( \frac{\delta S_0}{\delta \varphi ^a}\frac{%
\delta S_0}{\delta \varphi ^{*a}}+\frac 12\frac{\delta S_0}{\delta
B^{a\mu \nu }}\frac{\delta S_0}{\delta B_{\mu \nu
}^{*a}}+\frac{\delta S_0}{\delta
\eta ^{a\mu }}\frac{\delta S_0}{\delta \eta _\mu ^{*a}}+\frac{\delta S_0}{%
\delta \rho ^a}\frac{\delta S_0}{\delta \rho ^{*a}}\right)
\label{eme}
\end{equation}
\noindent The nilpotent BRST transformation $s$ on all fields and
antifields is

\begin{equation}
\begin{array}{ll}
s\varphi ^a=0\,, & s\varphi _a^{*}=-\partial ^2\varphi \,, \\ &
\\[1mm] sB_{\mu \nu }^a=\partial _\mu \eta _\nu ^a-\partial _\nu
\eta _\mu ^a\,, & sB_a^{*\mu \nu }=-\partial _\rho H_a^{\rho \mu
\nu }\,, \\ &  \\[1mm] s\eta _\mu ^a=\partial _\mu \rho ^a\,, &
s\eta _a^{*\mu }=\partial _\rho B_a^{*\rho \mu }\,, \\ &  \\[1mm]
s\rho ^a=0\,, & s\rho _a^{*}=-\partial _\mu \eta _a^{*\mu }\,. \\
&  \\[1mm] &
\end{array}
\label{brst}
\end{equation}

\noindent We shown in the table below, the canonical dimension and
the ghost number for all fields and antifields of the model
\begin{table}[hbt]
\centering

\begin{tabular}{|c|c|c|c|c|c|c|c|c|c|}
\hline
& $\Phi^{a}$ & $B^{a}_{\mu\nu}$ & $\eta^{a}_{\mu}$ & $\rho^{a}$ & $%
\Phi^{\ast a}$ & $B^{\ast a}_{\mu\nu}$ & $\eta^{\ast a}_{\mu}$ &
$\rho^{\ast a}$ &  \\ \hline $N_{g}$ & 0 & 0 & 1 & 2 & -1 & -1 &
-2 & -3 &  \\ \hline $dim$ & 1/2 & 1/2 & -1/2 & -3/2 & 5/2 & 5/2 &
7/2 & 9/2 &  \\ \hline
\end{tabular}


\caption[t1]{Ghost numbers and dimensions.} \label{gh1-number-dim}
\end{table}

Having the ghost number and dimension of all fields and antifields
at hand, we are now able to solve our problem using the consistent
deformation method. The action ~(\ref{s0}) will be deformed to a
new action $S$ in powers of the deformation parameters:
\begin{equation}  \label{def}
S=S_0+\sum_i g_i S_i+\sum_{i,j}g_i g_j S_{ij}+ \cdots~,
\end{equation}
where $S_i, S_{ij}..$ are local integrated polynomials with ghost
number zero and dimension bounded by three, and $g_i$ are the
deformed parameters with nonnegative mass dimension. The action
(\ref{def}) must satisfy the master equation
\begin{equation}  \label{cons}
(S,S)=0.
\end{equation}
Expanding the master equation (\ref{cons}) in powers of the
deformation parameters, we have
\begin{equation}  \label{order0}
(S_0,S_0)=0,
\end{equation}
\begin{equation}  \label{order1}
(S_0,S_i)=0,
\end{equation}
\begin{equation}  \label{order2}
2(S_0,S_{ij})+(S_i,S_j)=0.
\end{equation}
The equation (\ref{order0}) is the the master equation for the
$S_0$, and not gives any additional information. The equation
(\ref{order1}) tell us that $S_i$ has to be a BRST invariant under
(\ref{brst}). We must neglect BRST exacts, since this correspond
to fields redefinitions. The last equation (\ref{order2}) is
satisfied only if the antibracket $(S_i,S_j)$ is a trivial
cocycle.

Let us now construct all $S_i$ solution of equation
(\ref{order1}). First we focus our attention to terms that do not
deform the gauge symmetry, i.e, terms constructed with the fields
only. Due to trivial BRST transformation of $\varphi ^a$, the all
possible terms with this field are
\begin{equation}
S_1=\int \!\!d^3x~\left( \alpha _a\varphi _a\right) ,\quad
S_2=\int \!\!d^3x~\left( \alpha _{ab}\varphi _a\varphi _b\right) ,
\label{phionly12}
\end{equation}
\begin{equation}
S_3=\int \!\!d^3x~\left( \alpha _{abc}\varphi _a\varphi _b\varphi
_c\right) ,\quad S_4=\left( \int \!\!d^3x~\alpha _{abcd}\varphi
_a\varphi _b\varphi _c\varphi _d\right) ,  \label{phionly34}
\end{equation}
\begin{equation}
S_5=\int \!\!d^3x~\left( \alpha _{abcde}\varphi _a\varphi
_b\varphi _c\varphi _d\varphi _e\right) ,\quad S_6=\int
\!\!d^3x~\left( \alpha _{abcdef}\varphi _a\varphi _b\varphi
_c\varphi _d\varphi _e\varphi _f\right) ,  \label{phionly56}
\end{equation}
where $\alpha ^{\prime }s$ are parameters. The most general
invariant local integrable terms that can be constructed with
$B_{\mu \nu }^a$ and $\varphi ^a$ mixed are
\begin{equation}
S_7=\int \!\!d^3x~\left( m_{ab}\epsilon ^{\mu \nu \alpha }H_{a\mu
\nu \alpha }\varphi _b\right) ,\quad S_8=\int \!\!d^3x~\left(
m_{abc}\epsilon ^{\mu \nu \alpha }H_{a\mu \nu \alpha }\varphi
_b\varphi _c\right)   \label{mixed12}
\end{equation}
\begin{equation}
S_9=\int \!\!d^3x~\left( m_{abcd}\epsilon ^{\mu \nu \alpha
}H_{a\mu \nu \alpha }\varphi _b\varphi _c\varphi _d\right) ,
\label{mixed3}
\end{equation}
with $m_{ab}$ having dimension of mass, $m_{abc}$ of dimension $1/2$ and $%
m_{abcd}$ a dimensionless parameter.

Observing the table (\ref{gh1-number-dim}), it is easy to see that
it is impossible to construct invariant local integrated
polynomials with dimension bounded by three with the antifields.
This means that the algebra of the gauge symmetry is undeformed,
i.e., we do not have a nonabelian generalization of the action
(\ref{sa}), the only possibility being with an introduction of
extra fields or non-renormalizable couplings.

Let us now introduce a set of abelian vectorial gauge field in
order to implement the possible nonabelian generalization of
(\ref{sa}). We take the mass dimension of all vector equal to one,
to make an auxiliary character of those fields. The BRST
transformation are
\begin{equation}
sA_\mu ^a=\partial _\mu c^a,\quad \quad sc^a=0,  \label{nfield}
\end{equation}
where $c^a$ are the ghost for the abelian transformation of $A_\mu
^a$. We must add to the action (\ref{s0}) the corresponding
antifield action
\begin{equation}
S_{ant}^{\prime \prime }=\int \!\!d^3x~A_{a\mu }^{*}\partial ^\mu
c_a. \label{ant-new}
\end{equation}
The new antifields have the following BRST transformations
\begin{equation}
sA_\mu ^{*a}=0,\quad \quad sc^{*a}=\partial _\mu A^{*a\mu }
\label{brst-anti}
\end{equation}
We show in the table below the ghost number and dimension for new
fields (antifields)
\begin{table}[tbh]
\centering
\begin{tabular}{|c|c|c|c|c|} \hline
& $A_{\mu}^{a}$ & $c^{a}$ & $A^{\ast a}_{\mu}$ & $c^{\ast a}$ \\
\hline $N_{g}$ & 0 & 1 & -1 & -2 \\ \hline $dim$ & 1 & 0 & 2 & 3
\\ \hline
\end{tabular}
\caption[t2]{Ghost numbers and dimensions.} \label{gh2-number-dim}
\end{table}

The all possible invariant integrated local polynomials that can
be constructed with all fields and antifields are
\begin{equation}
S_{10}=g\int \!\!d^3x~f_{abc}\left( \varphi _a^{*}\varphi
_bc_c-\partial ^\mu \varphi _a\varphi _bA_{c\mu }\right)
,~S_{11}=\mu _{ab}\int \!\!d^3x~\left( A_{a\mu }^{*}\eta _b^\mu
-c_a^{*}\rho _b\right) ,
\end{equation}
\begin{equation}
S_{12}=h\int \!\!d^3x~k_{abc}\left( A_a^{*}A_b^\mu c_c-\frac
12c_a^{*}c_bc_c\right) ,
\end{equation}
where $g,h$ are dimensionless parameter, $\mu $ is a matrix with
dimension 3/2, and $f_{abc}(k_{abc})$ are dimensionless parameters
antisymmetric in its first(last) two indices. Now we perform the
calculation of the antibrackets $(S_i,S_j)$, with $i,j=1,2,\cdots
12$, in order to fit the second order consistency condition. As we
have already seen above, this
antibrackets must be a BRST exact. The antibrackets $(S_m,S_n)$, for $%
n,m=1,2,\dots ,9$ is identically zero, due to absence of antifields in $%
S_n,n=1,2,\dots ,9$. The antibracket $(S_{10},S_{10})$ is
\begin{eqnarray}
(S_{10},S_{10}) &=&g^2\int \!\!d^3x~f_{abc}f_{ab^{\prime
}c^{\prime }}\left( \varphi _{b^{\prime }}^{*}\varphi
_bc_cc_{c^{\prime }}+\varphi _{[b}\partial _\mu \varphi
_{b^{\prime }]}A_{c^{\prime }}^\mu c_c\right)   \nonumber \\
&&-\frac{g^2}2s\left( \int \!\!d^3x~f_{abc}f_{ab^{\prime
}c^{\prime }}\varphi _b\varphi _{b^{\prime }}A_{c\mu }A_{c^{\prime
}}^\mu \right) , \label{s10s10}
\end{eqnarray}
where, $\varphi _{[b}\partial _\mu \varphi _{b^{\prime }]}=\varphi
_b\partial _\mu \varphi _{b^{\prime }}-\varphi _{b^{\prime
}}\partial _\mu \varphi _b$. The first term in (\ref{s10s10}), is
not a BRST trivial and it could jeopardize the nonabelian
implementation. In order to circumvent this, we must have the
identification $hk_{abc}=gf_{abc}$, and $f_{abc}$ being the
structure constant of a Lie group. Therefore the $S_{10}$ and
$S_{12}$ are replaced by the sum
\begin{equation}
S_{10}^{\prime }=g\int \!\!d^3x~f^{abc}\left( \varphi ^{*a}\varphi
^bc^c-\partial ^\mu \varphi ^a\varphi ^bA_\mu ^c+A^{*a}A^{b\mu
}c^c-\frac 12c^{*a}c^bc^c\right) .  \label{S10l}
\end{equation}
It is easy to see that now $(S_{10}^{\prime },S_{10}^{\prime })$
is BRST trivial
\begin{equation}
(S_{10}^{\prime },S_{10}^{\prime })=-\frac{g^2}2s\left( \int
\!\!d^3x~f_{abc}f_{ab^{\prime }c^{\prime }}\varphi _b\varphi
_{b^{\prime }}A_{c\mu }A_{c^{\prime }}^\mu \right) .
\label{sl10sl10}
\end{equation}
The antibrackets $(S_{10}^{\prime },S_n)$, with $n=1,2,\dots ,6$,
gives us constraints for the parameters $\alpha $: $\alpha
_a=\alpha _{abc}=\alpha _{abcde}=0$, $\alpha _{ab}=a_1\delta
_{ab}$, $\alpha _{abcd}=a_2\delta _{ab}\delta _{cd}$, $\alpha
_{abcdef}=a_3\delta _{ab}\delta _{cd}\delta _{ef} $, i.e., only
the terms $\varphi ^2=\varphi _a\varphi _a$, $(\varphi ^2)^2$ and
$(\varphi ^2)^3$ are permitted. The last antibrackets reads
\[
(S_{11},S_{11})=0,
\]
\begin{eqnarray}
(S_{10}^{\prime },S_{11})=g\int \!\!d^3x~\!\!\!\! &&f_{abc}\mu
_{ab^{\prime }}\left( \rho _{b^{\prime }}\varphi _b^{*}\varphi
_c+\rho _{b^{\prime }}A_{b\mu }^{*}A_c^\mu -\rho _{b^{\prime
}}c_b^{*}c_c\right.   \nonumber \\ &&\left. -\eta _{b^{\prime
}}^\mu \partial _\mu \varphi _b\varphi _c-\eta _{b^{\prime }}^\mu
A_{b\mu }^{*}c_c\right) ,  \label{s11}
\end{eqnarray}
\[
(S_{10}^{\prime },S_7)=g\int \!\!d^3x~f_{abc}m_{b^{\prime
}a}\varepsilon _{\mu \nu \alpha }H_{b^{\prime }}^{\mu \nu \alpha
}\varphi _bc_c,
\]
\[
(S_{10}^{\prime },S_8)=g\int \!\!d^3x~f_{abc}(m_{b^{\prime
}c^{\prime }a}+m_{b^{\prime }ac^{\prime }})\varepsilon _{\mu \nu
\alpha }H_{b^{\prime }}^{\mu \nu \alpha }\varphi _b\varphi
_{c^{\prime }}c_c,
\]
\[
(S_{10}^{\prime },S_9)=g\int \!\!d^3x~f_{abc}(m_{b^{\prime
}c^{\prime }d^{\prime }a}+m_{b^{\prime }c^{\prime }ad^{\prime
}}+m_{b^{\prime }ac^{\prime }d^{\prime }})\varepsilon _{\mu \nu
\alpha }H_{b^{\prime }}^{\mu \nu \alpha }\varphi _b\varphi
_{c^{\prime }}\varphi _{d^{\prime }}c_c.
\]
The last four antibrackets are not BRST trivial, representing thus
an obstruction to the deformation of the master equation. The only
way to remedy this is setting $g=0$, or setting
$S_7=S_8=S_9=S_{11}=0$. In the case $g=0$ we have lost the
deformation of the abelian algebra, i.e, we have a set of abelian
fields not representing a nonabelian generalization of (\ref
{sa}). In the case in which $S_7=0$, we have lost the mass
generation of the model. We have thus proved that there are no
nonabelian generalization of the action (\ref{sa}), even with an
addition of an auxiliary vector gauge field.

\section{BRST and anti-BRST symmetry}

It is interesting to remark that the introduction of an one form
gauge connection $A$ is required to go further in the nonabelian
generalization of our model (\ref{sa}), although our original
abelian action (\ref{sa}) does not contain this field. Note that,
as pointed out by Thierry-Mieg and
Ne'eman \cite{thierry} for the nonabelian case, the field strenght for $B$ is%
\footnote{%
Here and in the rest of the paper, in order to handle BRST
transformations, we use differential forms formalism for
convenience.}

\begin{equation}
H=dB+\left[ A,B\right] \text{ }\equiv DB\text{ .}  \label{2.1}
\end{equation}
where $d=dx^\mu (\partial /\partial x^\mu )$ is the exterior
derivative.

Taking into account the no-go theorem shown in the previous
section, we must add an auxiliary field. Resorting to ref.
\cite{thierry}, we can define a new ${\cal H}$ given by

\begin{equation}
{\cal H}=dB+\left[ A,B\right] +\left[ F,C\right] \text{ },
\label{2.2}
\end{equation}
where $C$ is the one form auxiliary field required and
$F=dA+A\wedge A$.

The obstruction to the nonabelian generalization lies only on the
kinetic term for the antisymmetric field, but the topological term
must be conveniently redefined. So the nonabelian version of
(\ref{sa}) can be written as

\begin{equation}
\int_{M_3}Tr\left\{ \frac 12{\cal H}\wedge ^{*}{\cal H}+m{\cal
H}\wedge \varphi +\frac 12D\varphi \wedge ^{*}D\varphi \right\}
\text{ ,}  \label{2.3}
\end{equation}
where $*$ is the Hodge star operator.

The action above is invariant under the following transformations:

\begin{equation}
\delta A=-D\theta ,  \label{2.4}
\end{equation}

\begin{equation}
\delta \varphi =\left[ \theta ,\varphi \right] ,  \label{2.5}
\end{equation}

\begin{equation}
\delta B=D\Lambda +\left[ \theta ,B\right] \text{ ,}  \label{2.6}
\end{equation}
and

\begin{equation}
\delta C=\Lambda +\left[ \theta ,C\right]   \label{2.6a}
\end{equation}
where $\theta $ and $\Lambda $ are zero and one-form
transformation parameters respectively.

Here we shall use a formalism developed by Thierry-Mieg {\it et
al. }\cite {thierry,thierry2} in order to obtain the BRST\ and
anti-BRST tranformation rules. In general lines, we follow closely
the treatment of refs. \cite {thierry} or \cite{hwang}, since the
new object introduced here, namely the scalar field, does not
modify the approach.

The presence of a scalar field in topological invariants is not so
uncommon. A three-dimensional Yang-Mills topological action was
proposed by Baulieu and Grossman \cite{grossman} for magnetic
monopoles by gauge fixing the following topological invariant:

\begin{equation}
S_{top}=\int_{M_3}Tr\left\{ F\wedge D\varphi \right\} \text{ .}
\label{2.7}
\end{equation}

In the work of Thierry-Mieg and Ne'eman \cite{thierry}, a
geometrical BRST quantization scheme was developed where the base
space is extended to a fiber bundle space so that it contains
unphysical (fiber-gauge orbit) directions and physical
(space-time) directions. Using a double fiber bundle structure
Quiros {\it et al. }\cite{quiros} extended the principal fiber
bundle formalism in order to include anti-BRST symmetry. Basically
the procedure consists in extending the space-time to take into
account a pair of scalar anticommuting coordinates denoted by $y$
and $\overline{y}$ which correspond to coordinates in the
directions of the gauge group of the principal fiber bundle. Then
the so-called ''horizontality condition'' is imposed. This
condition enforces the curvature components containing vertical
(fiber) directions to vanish. So only the horizontal components of
physical curvature in the extended space survive.

Let us define the following form fields in the extended space and
valued in the Lie algebra ${\cal G}$ of the gauge group:

\begin{equation}
\widetilde{\varphi }=\varphi \text{ ,}  \label{2.8o}
\end{equation}

\begin{equation}
\widetilde{A}\equiv A_\mu dx^\mu +A_Ndy^N+A_{\overline{N}}d\overline{y}^{%
\overline{N}}\equiv A+\alpha +\overline{\alpha },  \label{2.8a}
\end{equation}

\begin{eqnarray}
\widetilde{B} &\equiv &\frac 12B_{\mu \nu }dx^\mu \wedge dx^\nu
+B_{\mu
N}dx^\mu \wedge dy^N+B_{\mu \overline{N}}dx^\mu \wedge d\overline{y}^{%
\overline{N}}+\frac 12B_{MN}dy^M\wedge dy^N  \nonumber \\
&&+B_{M\overline{N}}dy^M\wedge d\overline{y}^{\overline{N}}+\frac 12B_{%
\overline{M}\overline{N}}d\overline{y}^{\overline{M}}\wedge d\overline{y}^{%
\overline{N}}  \nonumber \\ &\equiv &B-\beta -\overline{\beta
}+\gamma +h+\overline{\gamma }, \label{2.8b}
\end{eqnarray}
and

\begin{equation}
\widetilde{C}\equiv C_\mu dx^\mu +C_Ndy^N+C_{\overline{N}}d\overline{y}^{%
\overline{N}}\equiv C+c+\overline{c}\text{ .}  \label{2.8c}
\end{equation}

Note that we identify the components in unphysical directions with
new
fields, namely, $\alpha $, $\beta $ and $c$ ($\overline{\alpha }$, $%
\overline{\beta }$ and $\overline{c}$) as anticommuting ghosts
(antighosts) and the commuting ghosts (antighost) $\gamma $ and
$h$ ( $\overline{\gamma }$ ).

The curvatures 2-form $\widetilde{F}$ and 3-form $\widetilde{{\cal
H}}$ in the fiber-bundle space are

\begin{equation}
\widetilde{F}\equiv \widetilde{d}\widetilde{A}+\widetilde{A}\wedge
\widetilde{A}  \label{2.9a}
\end{equation}
and

\begin{equation}
\widetilde{{\cal H}}\equiv \widetilde{d}\widetilde{B}+\left[ \widetilde{A},%
\widetilde{B}\right] +\left[ \widetilde{F},\widetilde{C}\right]
\text{ ,} \label{2.9b}
\end{equation}
where $\widetilde{d}=d+s+\overline{s}.$ The exterior derivatives
in the
gauge group directions are denoted by $s=dy^N(\partial /\partial y^N)$ and $%
\overline{s}=d\overline{y}^{\overline{N}}(\partial /\partial \overline{y}^{%
\overline{N}}).$

It is important to remark here that since we are focusing a mass
generation mechanism or, in other words, the action (\ref{2.3}),
the extra symmetries which appear in the pure topological model
have no room in the present discussion.

The horizontality condition, or equivalently, the Maurer-Cartan
equation for the field strenght $F$ can be written as

\begin{equation}
\widetilde{F}\equiv \widetilde{d}\widetilde{A}+\widetilde{A}\wedge
\widetilde{A}=F\text{ ,}  \label{2.10a}
\end{equation}
and for the 3-form ${\cal H}$ is

\begin{equation}
\widetilde{{\cal H}}\equiv \widetilde{d}\widetilde{B}+\left[ \widetilde{A},%
\widetilde{B}\right] +\left[ \widetilde{F},\widetilde{C}\right] ={\cal H}%
\text{ .}  \label{2.10b}
\end{equation}

Also we can impose the horizontality condition for the one form
$D\varphi $, which may be written as

\begin{equation}
\widetilde{D}\widetilde{\varphi }=\widetilde{d}\varphi +\left[ \widetilde{A}%
,\varphi \right] =D\varphi \text{ .}  \label{2.10c}
\end{equation}

By expanding both sides of (\ref{2.10a}) over the pairs of two
forms, one can obtain the following transformation rules:

\[
sA_\mu =D_\mu \alpha \text{ },\text{ }\overline{s}A_\mu =D_\mu \overline{%
\alpha }\text{ },
\]

\begin{equation}
s\alpha =-\alpha \wedge \alpha \text{ },\text{
}\overline{s}\overline{\alpha }=-\overline{\alpha }\wedge
\overline{\alpha }\text{ },  \label{2.11a}
\end{equation}

\[
s\overline{\alpha }+\overline{s}\alpha =-\alpha \wedge
\overline{\alpha }
\]

In order to close the algebra, we introduce an auxiliary scalar
commuting field $b$ valued in the Lie algebra ${\cal G}$ such that

\begin{equation}
s\overline{\alpha }=b\text{ ,}  \label{2.11b}
\end{equation}
and consequently

\begin{equation}
\overline{s}\alpha =-b-\overline{\alpha }\wedge \alpha \text{ , }\overline{s}%
b=-\overline{\alpha }\wedge b\text{ , }sb=0\text{ .}
\label{2.11c}
\end{equation}

On the other hand, expanding (\ref{2.10b}) over the basis of
3-forms yields

\[
sB_{\mu \nu }=-[\alpha ,B_{\mu \nu }]-D_{[\mu }\beta _{\nu
]}+[F_{\mu \nu },c]\text{ },\text{ }\overline{s}B_{\mu \nu
}=-[\overline{\alpha },B_{\mu \nu }]-D_{[\mu }\overline{\beta
}_{\nu ]}-[F_{\mu \nu },\overline{c}],
\]

\[
s\beta _\mu =-[\alpha ,\beta _\mu ]+D_\mu \gamma \text{ , }\overline{s}%
\overline{\beta }_\mu =-[\overline{\alpha },\overline{\beta }_\mu
]+D_\mu \overline{\gamma }
\]

\begin{equation}
s\overline{\beta }_\mu +\overline{s}\beta _\mu =-[\alpha ,\overline{\beta }%
_\mu ]-[\overline{\alpha },\beta _\mu ]+D_\mu h  \label{2.12}
\end{equation}

\[
s\gamma =-[\alpha ,\gamma ]\text{ },\text{ }\overline{s}\overline{\gamma }=-[%
\overline{\alpha },\overline{\gamma }]
\]

\[
\overline{s}\gamma +sh=-[\alpha ,h]-[\overline{\alpha },\gamma ]\text{ , }s%
\overline{\gamma }+\overline{s}h=-[\overline{\alpha },h]-[\alpha ,\overline{%
\gamma }]\text{ }
\]
Note that when we treat two odd forms, the $[$ $,$ $]$ must be
reading as an anticommutator.

The action of $s$ and $\overline{s}$ upon $c$, $\overline{c}$ and
$C$ is not defined in eq.(\ref{2.12}). However, the condition
(\ref{2.10b}) leads us to

\begin{equation}
\widetilde{B}+\widetilde{D}\widetilde{C}=B+DC\text{ .}
\label{2.13}
\end{equation}

The condition (\ref{2.13}) yields the BRST and anti-BRST
transformations for the auxiliary field $C$ and its ghosts $c$ and
$\overline{c}$:

\[
sC_\mu =-[\alpha ,C_\mu ]+D_\mu c+\beta _\mu \text{ , }\overline{s}C_\mu =-[%
\overline{\alpha },C_\mu ]+D_\mu \overline{c}+\overline{\beta
}_\mu \text{ ,}
\]

\begin{equation}
sc=-[\alpha ,c]-\gamma \text{ ,
}\overline{s}\overline{c}=-[\overline{\alpha
},\overline{c}]-\overline{\gamma }\text{ ,}  \label{2.14}
\end{equation}

\[
s\overline{c}+\overline{s}c=-[\overline{\alpha },c]-[\alpha ,\overline{c}]-h%
\text{ .}
\]

However, as usual, the action of $s$ and $\overline{s}$ on the
ghosts and
antighosts is not completely specified by eqs. (\ref{2.12}) and (\ref{2.14}%
). Therefore, a set of auxiliary fields is required, namely, a
commuting
vector field $t_\mu ,$ two anticommuting scalar fields $\omega $ and $%
\overline{\omega }$ and a commuting scalar field $n$. These fields
are used to solve eqs. (\ref{2.12}). Then, we get

\[
s\overline{\beta }_\mu =t_\mu \text{ , }\overline{s}\beta _\mu
=-t_\mu -[\alpha ,\overline{\beta }_\mu ]-[\overline{\alpha
},\beta _\mu ]+D_\mu h,
\]

\[
sh=\omega \text{ , }\overline{s}\gamma =-\omega -[\alpha ,h]-[\overline{%
\alpha },\gamma ],
\]

\[
s\overline{\gamma }=\overline{\omega }\text{ , }\overline{s}h=-\overline{%
\omega }-[\alpha ,\overline{\gamma }]-[\overline{\alpha },h],
\]

\begin{equation}
s\overline{c}=n\text{ , }\overline{s}c=-n-[\alpha ,\overline{c}]-[\overline{%
\alpha },c]-h\text{ ,}  \label{2.15}
\end{equation}

\[
st_\mu =s\omega =s\overline{\omega }=sn=0,
\]

\[
\overline{s}t_\mu =-[\overline{\alpha },t_\mu ]-[D_\mu \alpha ,\overline{%
\gamma }]-D_\mu \overline{\omega }-[\overline{\beta }_\mu ,t]\text{ , }%
\overline{s}n=-[\overline{\alpha },n]-[\overline{c},b]+\overline{\omega }%
\text{ , }
\]

\[
\overline{s}\omega =-[\overline{\alpha },\omega ]-[\alpha \alpha ,\overline{%
\gamma }]-[\alpha ,\overline{\omega }]-[h,b]\text{ , }\overline{s}\overline{%
\omega }=-[\overline{\alpha },\overline{\omega }]-[\overline{\gamma },b]%
\text{.}
\]
The nilpotency of the $s$ and $\overline{s}$ operators was used to
obtain the last eight relations.

Finally, by expanding (\ref{2.10c}), we obtain

\begin{equation}
s\varphi =[\alpha ,\varphi ]\text{ , }\overline{s}\varphi
=[\overline{\alpha },\varphi ]\text{ .}  \label{2.16}
\end{equation}

Therefore, a complete set of BRST and anti-BRST equations, namely,
eqs. (\ref {2.11a}-\ref{2.11c}), (\ref{2.14}-\ref{2.16}), and
(\ref{2.12}), associated with the classical symmetry
(\ref{2.4}-\ref{2.6}) was obtained.

In this confusion of auxiliary fields it is important to point out
the difference between the fields which do not belong to the
principal fiber
bundle expansion of the ''physical '' fields ($b,t_\mu ,n,\omega $ and $%
\overline{\omega }$) (introduced in order to complete the
BRST/anti-BRST algebra) and the auxiliary one form field $C$
introduced in order to overcome the obstruction to the nonabelian
generalization. Note that here the  {\it a priori }introduction of
the auxiliary field $C$, was necessary in order to fix the BRST
and anti-BRST transformation rules. This suggests an interesting
and remarkable connection between the technique of consistent
deformation and the horizontality condition.

It is worth to mention that in $D=3$ a purely topological model
involving a
mixed Chern-Simons term ( two different one form fields) and the term $%
B\wedge D\varphi $ was discussed in ref. \cite{delcima}, and its
finiteness was proved in the framework of algebraic
renormalization.

We end up this section by observing that the obstruction to
nonabelian generalization of the four-dimensional BF model,
namely, the existence of
the constraint $[F,^{*}H]=0,$ appears in the context of our model as $%
[F,^{*}H-m\varphi ]=0$, as can be seen from the equations of
motion of the action (\ref{2.3}), considered in the absence of the
auxiliary field.

\section{\bf Nonabelian Topological Mass Generation}

The simplest scenario to study mass generation is to consider the
equations of motion of the action (\ref{2.3}). For convenience, we
define a new one form field as

\begin{equation}
K\equiv D\varphi  \label{3.0}
\end{equation}

Therefore, the equations of motion can be written as

\begin{equation}
D^{*}{\cal H}=mK  \label{3.1}
\end{equation}
and

\begin{equation}
D^{*}K=-m{\cal H}.  \label{3.2}
\end{equation}

Equations (\ref{3.1}) and (\ref{3.2}) can be combined into the
following second order equations:

\begin{equation}
\left( D^{*}D^{*}+m^2\right) {\cal H}=0  \label{3.3}
\end{equation}

\begin{equation}
\left( D^{*}D^{*}+m^2\right) K=0,  \label{3.4}
\end{equation}

Considering only linear terms for the fields, we get

\begin{equation}
\left( d^{*}d^{*}+m^2\right) H=0,  \label{3.5}
\end{equation}

\begin{equation}
\left( d^{*}d^{*}+m^2\right) d\varphi =0.  \label{3.6}
\end{equation}
which are similar to the eqs. (\ref{gmt061}) and (\ref{gmt062}),
and exhibit mass generation for $H$ and $\varphi .$

On the other hand, by looking to the pole structures of the
propagators of the model, mass generation can also be established.
In order to obtain them, we use the action (\ref{2.3}) added with
convenient gauge fixing terms, namely

\begin{eqnarray}
S_T &=&\int_{M_3}Tr\left\{ \frac 12{\cal H}\wedge ^{*}{\cal H}+m{\cal H}%
\wedge \varphi +\frac 12D\varphi \wedge ^{*}D\varphi +\right.
\nonumber \\ &&\left. {\cal J}\text{ }^{*}B+j\text{ }^{*}\varphi
+J\text{ }^{*}M+J_p\text{ }^{*}p+p^{*}dM+M\text{ }^{*}dB\right\} ,
\label{3.7}
\end{eqnarray}
where ${\cal J}$, $J$ , $J_p$ and $j$ are currents related to the
fields $B,$ $M,$ $p$ and $\varphi $ respectively, which generate
propagators in the path integral formulation. The auxiliary fields
$M$ and $p$ are introduced in order to implement the Landau gauge
fixing.

Therefore, the tree-level effective propagators for the
Kalb-Ramond and scalar fields are

\begin{equation}
<\varphi \varphi >_{a,b}=-\frac{\delta _{ab}}{p^2-m^2}  \label{3.8}
\end{equation}
and

\begin{equation}
<BB>_{a\mu \nu ,b\rho \sigma }=\frac{\delta _{ab}}{p^2-m^2}\left[ g_{\mu
[\rho }g_{\sigma ]\nu }-\frac{g_{\mu [\rho }p_{\sigma ]}p_\nu }{p^2}+\frac{%
g_{\nu [\rho }p_{\sigma ]}p_\mu }{p^2}\right] ,  \label{3.9}
\end{equation}
where $a$ and $b$ are group indices, and $\mu ,\nu ,\rho $ and
$\sigma $ are space-time indices.

It is interesting to note that, here, the gauge field $B$ ''eats
'' the scalar field (not a Higgs field, however) and acquires a
longitudinal degree of freedom and a mass. The inverse process is
possible too.

\section{Conclusions}

In this work we have succeeded in extending a tridimensional
abelian topological model to the nonabelian case. The model
considered here couples a second rank antisymmetric tensor field
and a scalar field in a topological way. Initially we use the
method of consistent deformations to analyze upon what conditions
this generalization can be implemented. Then we have shown that if
we require power-counting renormalizable couplings and the same
field content, the nonabelian extension is forbidden.

We overcome this obstruction by introduction of two new fields in
the model in order to obtain the pursued nonabelian version. One
field is a one form gauge connection ( $A$ ) which allows us to
define a Yang-Mills covariant derivative. The other auxiliary
field ($C$ ) is a vectorial one, which is required in order to
resolve the constraint that prevents the correct
nonabelianization.

A formal framework to consider the introduction of these fields
and the consequent new symmetries, is furnished by BRST and
anti-BRST transformation rules, which are obtained using the
horizontality condition. Although quite similar to other
topological models, it is worth to mention that, in this case, we
have constructed transformation rules for the Kalb-Ramond field,
for two one form fields and for a scalar field.

Finally, the topological mass generation mechanism for an abelian
model found out in a previous paper was extended for the
nonabelian case, and we end up with an effective theory describing
massive Kalb-Ramond gauge fields in $D=3$ space-time.

We conclude mentioning the possible relevance of the present
discussion to string theory. Indeed, the Kalb-Ramond field couples
directly to the worldsheet of strings, and bosonic string
condensation into the vacuum realize the Higgs mechanism to the
Kalb-Ramond gauge field \cite{rey}. Therefore an alternative
scenario to give mass to the Kalb-Ramond field in the context of
strings may be an interesting continuation of our present results.

\vspace{0.3in} \centerline{\bf ACKNOWLEDGMENTS}

We would like to thank O. S. Ventura for helpful discussions. This
work was supported in part by Conselho Nacional de Desenvolvimento
Cient\'{\i }fico e Tecnol\'{o}gico-CNPq and Funda\c{c}\~{a}o
Cearense de Amparo \`{a} Pesquisa-FUNCAP.

\end{document}